\documentclass[reprint, superscriptaddress,amsmath,amssymb,aps,prl]{revtex4-1}
\usepackage{graphicx}
\usepackage{xcolor}
\usepackage{physics}
\usepackage{upgreek}

\begin{document}

\title{Coherent control for qubit state readout}

\author{Conrad Roman}
\author{Anthony Ransford}
\author{Michael Ip}
\author{Wesley C. Campbell}
\affiliation{University of California Los Angeles}

\date{\today}

\begin{abstract}
  Short pulses from mode-locked lasers can produce background-free atomic fluorescence by allowing temporal separation of the prompt incidental scatter from the subsequent atomic emission. We use this to improve quantum state detection of optical-frequency and electron-shelved trapped ion qubits by more than 2 orders of magnitude. For direct detection of qubits defined on atomic hyperfine structure, however, the large bandwidth of short pulses is greater than the hyperfine splitting, and repeated excitation is not qubit state selective. Here, we show that the state resolution needed for projective quantum measurement of hyperfine qubits can be recovered by applying techniques from coherent control to the orbiting valence electron of the queried ion. We demonstrate electron wavepacket interference to allow readout of the original qubit state using broadband pulses, even in the presence of large amounts of background laser scatter.

\end{abstract}

\maketitle

An essential capability of a quantum information processor is the ability to read out the result of the computation \cite{DiVincenzoCriteria}. Since these devices are susceptible to information corruption, it is also likely some form of quantum error correction will be required to perform even moderately lengthy computations, and repeated qubit state detections will be needed during operation \cite{Gottesman2009AnIntroduction}. The  primary metrics for evaluating the quality of qubit measurement for these tasks, therefore, are the fidelity \cite{Christensen2019High} and speed \cite{Noek2013Optical}, both of which frequently compare poorly to gate operations.

State readout in trapped ion processors is typically performed by illuminating ions with a continuous-wave (cw) laser resonant with a cycling transition that contains one qubit state (the so-called ``bright state'') and not the other (the ``dark state'') \cite{Nagourney1986Shelved,Bergquist1986Observation,Sauter1986Observation}. When an ion qubit is in the bright state, laser-induced fluorescence (LIF) photons are collected by high-NA imaging optics, spatially filtered from the incidental laser scatter, and counted using a photomultiplier tube (PMT). A dark state ion qubit, on the other hand, will not fluoresce and with ideal spatial filtering of the excitation light, no photons will be counted. For atomic hyperfine qubits, the state interrogation time is typically limited by off-resonant absorption of photons that mix the qubit basis states \cite{Acton2006}. Higher efficiency collection of the fluorescence typically improves both the fidelity and speed of the readout \cite{VanDevender2010Efficient,Merrill2011Demonstration,Noek2013Optical,Ghadimi2017Scalable}.

As trapped ion quantum processors continue to miniaturize \cite{Mehta2014Ion,Guise2015BallGrid,Maunz2015Characterization} and integrate photonic elements onto the trap chip \cite{Mehta2016Integrated,Kielpinski2016Integrated}, the ions are held increasingly close to surfaces that scatter excitation light toward the detector, and background scatter, as opposed to collection efficiency, can become the limiting factor in readout speed and fidelity \cite{Crain2019Highspeed}. In particular, scalable systems incorporating chip-integrated non-imaging single-photon detectors \cite{Eltony2013Transparent,Slichter2017UVsensitive} can achieve very high detection efficiency (particularly as the ion is moved closer to the detector), but must rely on other mechanisms for rejecting background scatter.

In this article, we report the use of a mode-locked laser to achieve qubit state detection that is nearly background free
by counting photons only in the dark time between the excitation pulses.  We first apply this method to detection of electron-shelved \cite{dehmelt:1975} hyperfine qubits, where the resolution required for state selectivity ($\approx\!170\mbox{ THz}$) is much larger than the bandwidth of the pulse ($\approx \! 50 \mbox{ GHz}$).  Temporal rejection of background scatter eliminates it as a constraint \cite{Nagourney1997Signaltonoise} and allows us to achieve a factor of 100 improvement in the signal to background ratio for single-shot detection with an average fidelity of $\mathcal{F}=0.9993^{+0.0003}_{-0.0006}$ in the presence of $\approx\!45\;000$ background counts per second. For un-shelved hyperfine qubits, however, the pulse bandwidth far exceeds the frequency resolution required for state selectivity ($\approx 2\mbox{ GHz}$).  Here, we bring together techniques from quantum information science and coherent control, and use Ramsey spectroscopy to gain a sub-femtosecond view of the complex motion of the valence electron orbiting this single atom between excitation pulses.  This unique, time-domain picture of the inner workings of a quantum bit allows us to discern and control the microscopic dynamics of the atom's electron to demonstrate direct detection of the qubit state by suppressing unwanted excitations through destructive interference of electron wavepackets.  This demonstration suggests that further application of coherent control techniques to quantum information science may provide solutions to some of the challenges facing the creation of large-scale, programmable quantum processors.

\begin{figure*}[th!]
\includegraphics[width=7in]{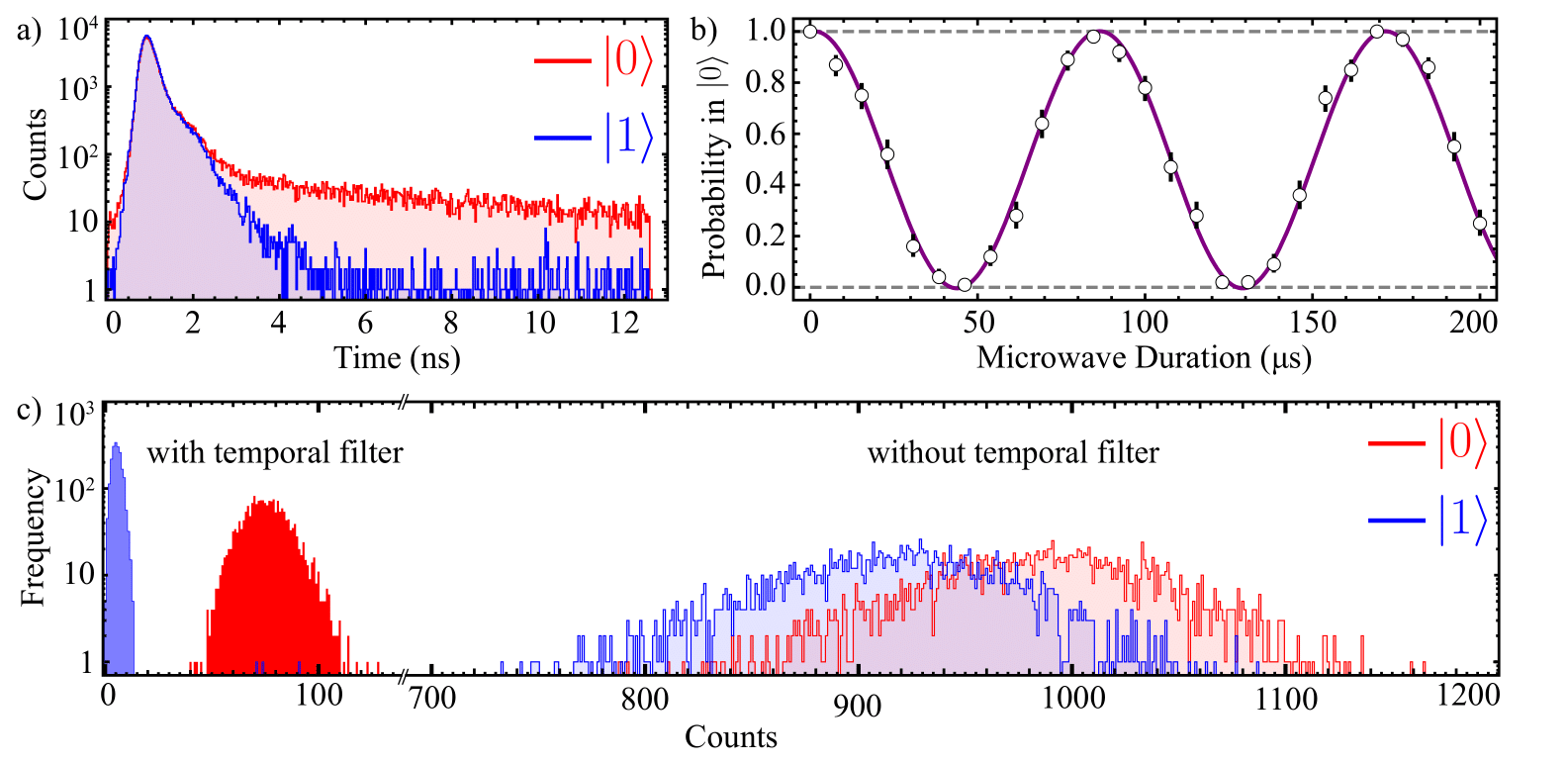}
\caption{\label{ShelvedHistograms} Single-shot background-free state detection of a trapped ion qubit through mode-locked excitation and temporal filtering for 20 ms illumination.  (a) Time trace of the detected photons for a shelved hyperfine qubit initially prepared in the bright (red) or dark (blue) qubit state.  (b) High fidelity detection of microwave-driven Rabi flops of the hyperfine qubit enabled by temporal rejection of background scatter.  (c) State detection histograms for the shelved hyperfine qubit shown with (left) and without (right) temporal filtering.  With temporal filtering, the fidelity is no longer limited by background scatter, and the remaining error is from imperfect shelving.}
\end{figure*}

\begin{figure*}[th!]
\includegraphics[width=7in]{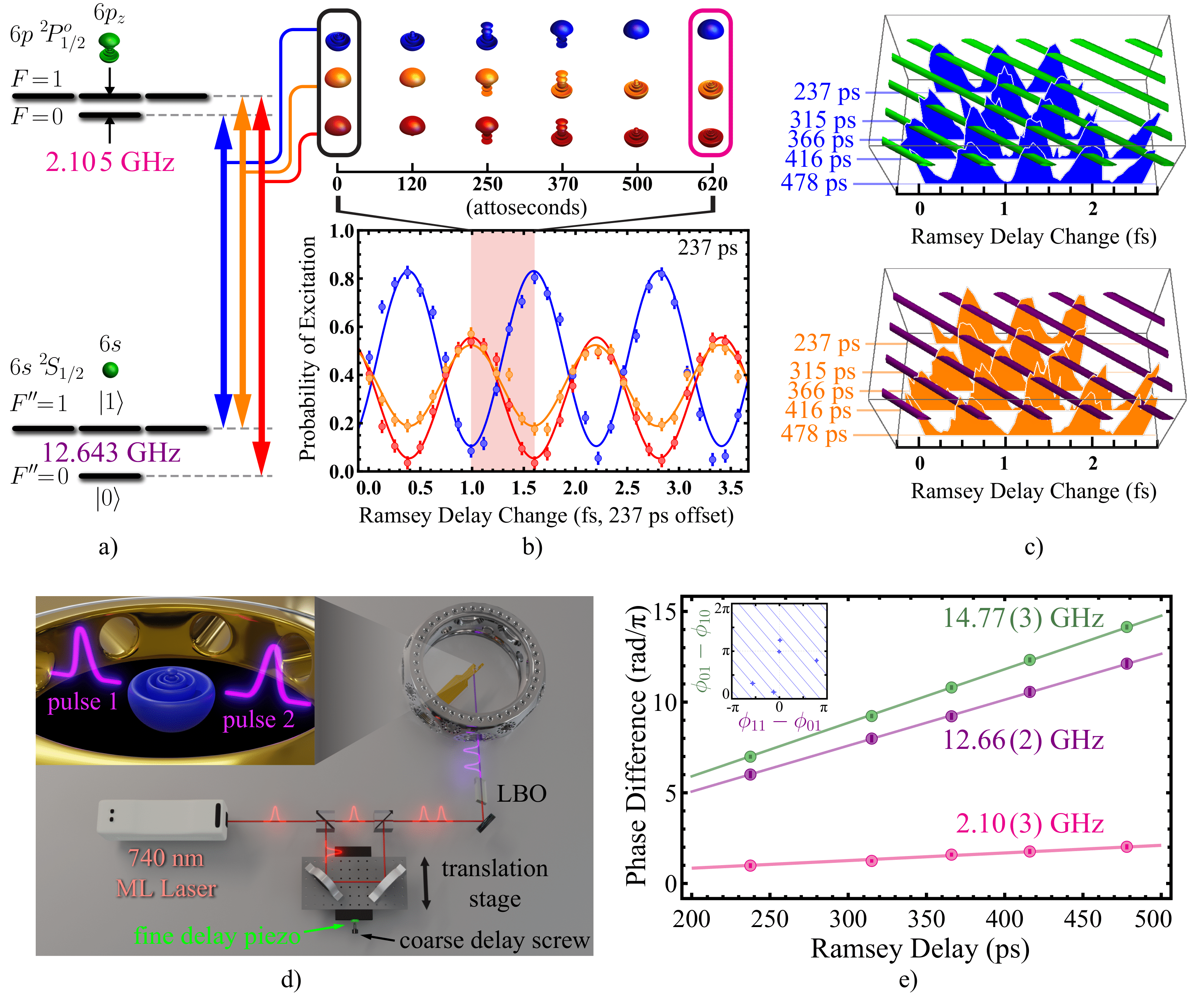}
\caption{\label{TriRamseyFig}Phase synchronization of electron wavepackets for qubit state detection.  a) Atomic structure of ${}^{171}\mbox{Yb}^+$.  The clock state qubit in the ${}^2S_{1/2}$ ground state is labeled with $\ket{1}$ and $\ket{0}$.  For projective measurement of the qubit, the transition indicated by the blue arrow must be driven while those indicated by the red and orange arrows should not.  b)  The three 2-level systems associated with these transitions will begin oscillating in phase right after a single, broadband excitation pulse.  After a delay time of $237 \mbox{ ps}$, the wavepackets excited on the red and orange transitions will be $\approx\!\pi$ out of phase from the blue transition, shown schematically as hydrogenic wavefunctions with the measured phase (top).  A second laser pulse can then be made to excite electron wavepackets that interfere constructively (destructively) with the blue (orange and red) wavepackets by setting its timing and phase to the top of a Ramsey fringe for the desired transition (outlined in magenta).  The excitation from this two-pulse sequence shows Ramsey fringes that can be measured independently (lower). c)  Ramsey fringes taken at a series of different coarse ($\mathcal{O}(50\mbox{ ps})$) delays, displaced vertically by the amount of this delay, to make time-domain Loomis-Wood diagrams.  Green (Purple) lines show the alignment of the $\ket{1} \! \leftrightarrow \! \ket{{}^2P_{1/2}^o;F\!=\!0}$ ($\ket{1} \! \leftrightarrow \! \ket{{}^2P_{1/2}^o;F\!=\!1}$) Ramsey fringes over $\approx \!10^5$ oscillations as referenced to the phase of the $\ket{0} \! \leftrightarrow \! \ket{{}^2P_{1/2}^o;F\!=\!1}$ electron wavepacket oscillations.  d) Delay stage for generating the pulse pairs of the Ramsey sequence.  A mechanical stage is used for coarse delay changes while a piezo device controls fine delays. e)  Following the relative phases of the Ramsey fringes over a substantial change in the coarse delay and extracting the slopes of linear fits allows Fourier-transform ultraviolet (FTUV) spectroscopy of the hyperfine structure in a single atom with a broadband pulse. }
\end{figure*}

The qubit in this work is hosted by a single $^{171}\text{Yb}^+$ ion confined in a radio frequency Paul trap with oblate spheroidal symmetry \cite{Yoshimura2015creation}.
The hyperfine clock-state qubit is defined in the ground $^2S_{1/2}$ state by the levels $\ket{0} \equiv \ket{F^{\prime\prime} \!= \!0, m_{F^{\prime\prime}} = 0}$ and  $\ket{1} \equiv \ket{F^{\prime\prime}\! =\! 1, m_{F^{\prime\prime}} = 0}$ \cite{Olmschenk2007Manipulation}, shown in Fig.~\ref{TriRamseyFig}a. For each experiment, after cw Doppler cooling of the ion's motion, qubit state preparation is performed by optical pumping to the $\ket{0}$ state via resonant, cw excitation to the excited $|{}^2P_{1/2}^o$; $F\! =\! 1\rangle$ manifold. Single-qubit gates for population transfer between qubit states are provided by resonant microwaves delivered via a microwave antenna.

In the typical, cw laser driven $I\!=\!\frac{1}{2}$ qubit state readout scheme \cite{Acton2006}, $\ket{1}$ acts as the bright state; a single-frequency cw laser resonant with $|{}^2S_{1/2};F^{\prime\prime} \! = \!1 \rangle \!\rightarrow \!|{}^2P^{o}_{1/2};F\! = \!0 \rangle$ is applied and excited atoms spontaneously decay back to $|{}^2S_{1/2};F^{\prime\prime} \! = \!1 \rangle$ with high probability \cite{Olmschenk2007Manipulation}.  This cycle is driven until the probability that off-resonant excitation mixes the qubit states limits the fidelity.  In contrast, for electron-shelved detection \cite{dehmelt:1975}, $\ket{0}$ plays the role of bright state; population in $\ket{1}$ is transferred (``shelved'') to a metastable state and photon collection during subsequent cw Doppler cooling is used to discern whether the ion was shelved.  For shelving of ${}^{171}\mathrm{Yb}^+$ hyperfine qubits \cite{RansfordShelving}, we apply a laser at $411\mbox{ nm}$ to transfer population from $\ket{1}$ to the effectively stable ${}^2F_{7/2}^o$ state \textit{via} the ${}^2D_{5/2}$ state. Subsequent illumination at $369.5\mbox{ nm}$ and $935.2\mbox{ nm}$ reveals essentially unlimited LIF for an ion found in $\ket{0}$, whereas an ion initially in $\ket{1}$ has been shelved and is dark.

For the pulsed qubit state detection in this work, excitation of the bright state population to ${}^2P^{o}_{1/2}$ is driven by a frequency doubled mode-locked (ML) laser generating near-transform-limited $\approx\!10\mbox{ ps}$ pulses at $369.5\mbox{ nm}$ with a repetition rate of $f_\mathrm{r}=79.5\mbox{ MHz}$. Because the laser-excited ${}^2P^{o}_{1/2}$ state has a lifetime of $\tau \! = \!8.12\mbox{ ns}$ \cite{Olmschenk2009Measurement}, an atom excited by a pulse from the ML laser will emit a photon with high (78$\%$) probability before the next pulse arrives. Nonetheless, multi-pulse coherence can still lead to strong comb tooth effects \cite{Ip2018phonon}, and the positions of the laser's optical frequency comb teeth in this work are held far from any resonances by an intra-cavity piezo-mounted mirror. Pulse energies at the ion are typically $\approx 0.25\mbox{ pJ}$, corresponding to a rotation of the atomic $^2S_{1/2} \! \leftrightarrow \! {}^2P_{1/2}^o$ Bloch vector of $\;\theta \approx 0.05\pi $. Ion fluorescence at $369.5\mbox{ nm}$ is collected by an objective lens and registered by a PMT through a $369\mbox{ nm}$ bandpass filter. The PMT output is monitored by either an FPGA-based photon time tagger ($\approx 10\mbox{ ns}$ resolution \cite{Pruttivarasin2015compact}) or a fast, time correlated single photon counter (TCSPC, $\approx 25\mbox{ ps}$ resolution \footnote{PicoQuant TimeHarp 260P}).

For shelved and optical-frequency qubits, the population in ${}^2F_{7/2}^o$ is well separated from the ground and excited states (by $> 100\mbox{ THz}$) and sits far outside the spectral bandwidth of the excitation pulse.  Figure~\ref{ShelvedHistograms}a shows a time trace of the collected photons from many repeated excitations of a single ion initially prepared in the dark ($|1\rangle$, blue trace) and bright ($|0\rangle$, red trace) hyperfine qubit states, after attempted shelving.  The prompt peak near $t= 1 \mbox{ ns}$ is caused by incidental laser scatter, and is extinguished below the level of dark counts ($40 \mbox{ s}^{-1}$) in about $4 \mbox{ ns}$. The observed pulse width of greater than 100 ps is due to the transit time spread of photo-electrons in the PMT. The window from $\approx 3-12.5 \mbox{ ns}$, in which an initially excited ion will emit a photon with roughly 50\% probability, allows fluorescence detection that is essentially free of corruption by background scatter.

By separately recording the number of collected photons for an ion initially prepared in dark state and the bright state for many repetitions, histograms of the two results are constructed (Fig.~\ref{ShelvedHistograms}c).  Well-separated histograms are required for single-shot state detection, where a threshold is typically chosen between the two distributions to discriminate the two states.  However, the total collected light, shown on the right side of Fig.~\ref{ShelvedHistograms}c, shows substantial overlap of the two histograms and poor state discrimination due to incidental scatter from the excitation laser, yielding an average single-shot detection fidelity of $\mathcal{F}=0.76\pm 0.01$ using a threshold.

In contrast, by repeating the same experiment while rejecting counts outside the background-free window, shown by the solid histograms on the left side of Fig.~\ref{ShelvedHistograms}c, essentially all of the substantial technical scatter caused by the mode-locked laser can be rejected.  This technique achieves an average single-shot detection fidelity of $\mathcal{F} = 0.9993^{+0.0003}_{-0.0006}$, and can be used, for instance, to observe high-contrast Rabi flopping of the qubit (Fig.~\ref{ShelvedHistograms}b).  As a result, despite the high amount of scatter introduced by using a mode-locked laser for qubit state detection, temporal rejection of background counts effectively eliminates the background scatter as a concern for detection fidelity. The remaining errors (the three blue counts in the solid red histogram on the left side of Fig.~\ref{ShelvedHistograms}c) are consistent with the bright state histogram and are therefore likely due to incomplete shelving, not poor manifold discrimination.  Further improvements in shelving \cite{RansfordShelving} can be used to reduce this infidelity, and this type of error will be entirely absent for optical-frequency qubits.

Given the improvement that this technique can provide for scatter-dominated environments, it is well suited for chip-based systems with integrated detectors, where high solid angle can be achieved, but scatter cannot be well filtered spatially.  In particular, integrated superconducting nanowire single photon detectors, which feature high efficiency at relevant wavelengths, low dark counts, high maximum count rates, and fast timing resolution, are being integrated into next-generation ion systems \cite{Slichter2017UVsensitive,Wollman2017UVSuperconducting,Crain2019Highspeed}.  With an estimated fractional solid angle coverage of 5\% (taken from the middle of the predicted 3\%-8\% range for future integrated SNSPDs \cite{Slichter2017UVsensitive}) and a system detection efficiency of 60\%, an optical-frequency qubit (or pre-shelved hyperfine qubit) in the bright state will able produce an average of 10 detector counts in $\approx\!5\mbox{ }\upmu\mbox{s}$.  Even if complete temporal filtering of the incidental scatter requires discarding half of these LIF counts, the achievable fidelity is $>99\%$ in only $5\mbox{ }\upmu\mbox{s}$.

For (un-shelved) hyperfine qubits, which may offer an improvement in aggregate detection speed compared to schemes that require shelving and deshelving steps, the pulse bandwidth ($\approx 50\mbox{ GHz}$) encapsulates the entire hyperfine substructure of the cycling transition (which has splittings from $2.105\mbox{ GHz}$ to $14.748\mbox{ GHz}$, see Fig.~\ref{TriRamseyFig}a). Excitation by a single pulse (followed by spontaneous emission) will therefore not only occur for both qubit states, but will also not preserve the total angular momentum $F^{\prime\prime}$ of the initial state.  For state discrimination, due to finite collection efficiency, the bright state should ideally spontaneously emit many photons on average, while the dark state should not.  As shown in Fig.~\ref{TriRamseyFig}a, this means the $|{}^2S_{1/2};F^{\prime\prime} \! = \! 1 \rangle \rightarrow |{}^2P_{1/2}^o;F \! = \! 0 \rangle$ transition (blue arrow), which will preserve $F^{\prime\prime}$, should be driven while excitations on the other two allowed pathways ($|{}^2S_{1/2};F^{\prime\prime} \! = \! 1 \rangle \rightarrow |{}^2P_{1/2}^o;F \! = \! 1 \rangle$ and $|{}^2S_{1/2};F^{\prime\prime} \! = \! 0 \rangle \rightarrow |{}^2P_{1/2}^o;F \! = \! 1 \rangle$, indicated by the orange and red arrows) should be strongly suppressed.  To employ the temporal background rejection technique demonstrated above for direct hyperfine qubit detection therefore requires a means to drive hyperfine-selective optical cycling with the broadband laser.

The route to achieving state selectivity can be found by examining the time dynamics of each excitation, which can be probed through Ramsey (\textit{i.e.}~coherent pump-probe) spectroscopy.  To implement a Ramsey experiment, a single pulse from the ML laser is split into two pulses by an imbalanced Mach-Zehnder interferometer (Fig.~\ref{TriRamseyFig}d). The delay of the second pulse with respect to the first is controlled with sub-fs precision by a piezo actuator.

Figure~\ref{TriRamseyFig}b (lower) shows three sets of Ramsey fringes obtained from the same ion that allow us to see, with sub-femtosecond resolution, the harmonic motion of the electron wavepacket associated with each of the three, approximately-isolated, two-level systems (TLSs) excited by the first pulse (for details of how these fringes were obtained, see supplemental information \cite{SupplementalMaterials}).  This simplified 3-TLS model can be used to discern the essential features of the complex dynamics of this multi-level quantum system to engineer a strategy to regain state selectivity. On a microscopic level, the first pulse (or Ramsey zone) excites an electron wavepacket that oscillates at $811\mbox{ THz}$ in the delay between pulses, illustrated schematically as $n\!\!\,=\,\!\!6$ hydrogenic orbital superpositions in Fig.~\ref{TriRamseyFig}b (upper).  While the wavepackets corresponding to each of the three TLSs will initially oscillate in phase with one another, the $\mathcal{O}(10\mbox{ ppm})$ difference between their oscillation frequencies will cause them to slip in relative phase as the delay increases.  Figure~\ref{TriRamseyFig}b shows the oscillations of each wavepacket after $\approx 237 \mbox{ ps}$ of Ramsey delay, at which point the two undesired excitations (shown in orange and red) are approximately $\pi$ out of phase with the desired transition (blue).  Sinusoidal fits of these data (solid curves) give relative phase differences of $\Delta \phi_\mathrm{orange, blue} = (0.99 \pm 0.04)\pi$ and $\Delta \phi_\mathrm{red, blue} = (0.98 \pm 0.04)\pi$.

Ramsey fringes taken at a series of different coarse (tens of picoseconds) delays can be coherently combined only if the number of full $(2\pi)$ phase windings between them can be determined.  While this is challenging for the optical frequencies here, by choosing one of the three fringe sets to act as a clock reference, phase \emph{differences} between sets will oscillate slower as a function of delay. Figure \ref{TriRamseyFig}c shows the blue (upper) and orange (lower) fringes, translated to set the peak of the red fringe at $t\!=\!0$, displayed displaced vertically by their coarse delay in the manner of a Loomis-Wood diagram \cite{LoomisWood} (see, e.g., \cite{Kisiel2005Rotational} for a modern example of the Loomis-Wood technique of molecular spectroscopy).  This diagrammatic approach allows patterns in the peaks to be discerned, which are shown as diagonal stripes in Fig.~\ref{TriRamseyFig}c.  The slope of these lines allow us to follow the phases of the fringes over a range of coarse delays, which amounts to performing Fourier-transform ultraviolet (FTUV) spectroscopy of the qubit atom.  Figure \ref{TriRamseyFig}e shows the results of the FTUV spectroscopy, which accurately reproduce the known atomic structure essentially to the level of the natural linewidth ($1/\tau\! = \!2 \pi \times 19.6 \mbox{ MHz}$), roughly 3 orders of magnitude narrower than the pulse bandwidth, and $50\times$ narrower than the Fourier width of the $500\mbox{ ps}$ measurement window \cite{Lomsadze1389}.

\begin{figure}[h]
  \begin{centering}
    \includegraphics[width = 0.9\linewidth]{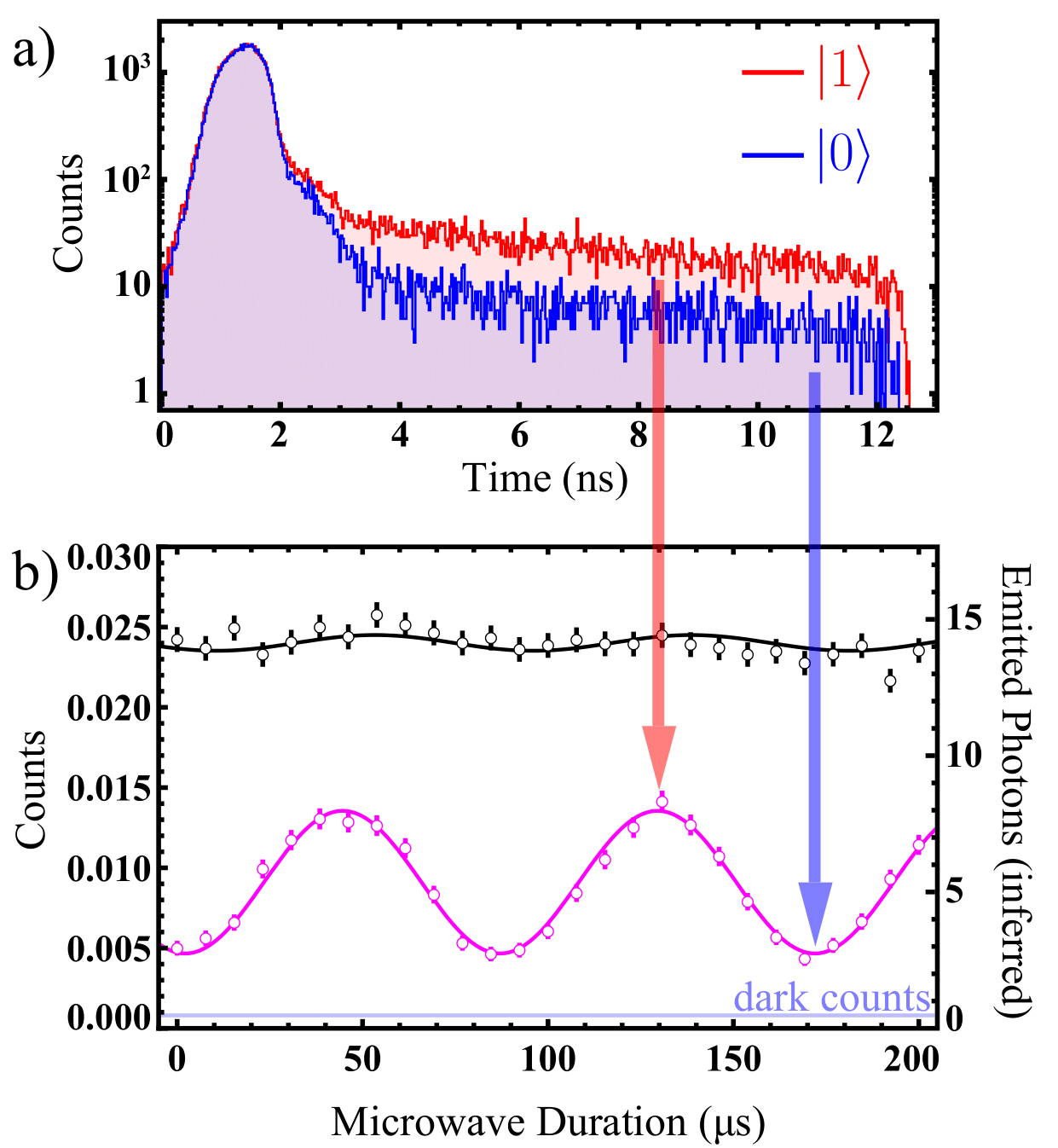}
    \caption{Direct detection of the hyperfine qubit with a $25\mbox{ }\upmu\mbox{s}$ train of split broadband pulses.  a) Arrival times of counts for direct detection of a qubit initially prepared in each eigenstate.  b)  When the Ramsey delay is chosen for constructive interference of the desired transition (outlined in magenta in Fig.~\ref{TriRamseyFig}b (upper)), Rabi flopping from a resonant microwave drive is apparent (magenta).  By moving the delay half a fringe from this point (outlined in black in Fig.~\ref{TriRamseyFig}b (upper)), only the two undesired transitions are driven, and the state discrimination of the collected fluorescence is very weak (black).  In the latter case, the total fluorescence rate is higher since the two undesired transitions act as hyperfine repumps for one another.}
    \label{fig:UnshelvedResults}
  \end{centering}
\end{figure}

By timing the second Ramsey pulse to be centered at the top of the fringe for the desired transition (outlined in magenta at the top of Fig.~\ref{TriRamseyFig}b), constructive interference of the transition amplitude driven by the second pulse with the coherent superposition from the first will drive the desired transition, while the amplitudes of the wavepackets for the two undesired transitions will destructively interfere with the previously excited wavepackets, and all of the population will be returned to the ground state.
In the limit of small pulse area, the ratio of the scattering rate of the desired transitions to the two undesired rates is predicted to be $\kappa \equiv \frac{\Gamma_\mathrm{d}}{\sum_u \Gamma_u} \approx \frac{4 \Delta}{\gamma}-\frac{1}{2}$, where $\gamma$ is the natural linewidth and $1/2\Delta$ is the delay time between the two pulses.  For ${}^{171}\mathrm{Yb}^+$, we calculate $\kappa \approx 68$ photons can theoretically be spontaneously emitted on average before mixing the qubit.

To evaluate the validity of the 3-TLS model and verify that the temporally-filtered LIF from this two-pulse excitation can be used to perform a direct, projective measurement of the qubit state, a microwave rotation of the qubit is again inserted between state preparation and (now double-pulsed) state detection, shown in Fig.~\ref{fig:UnshelvedResults}b.  A detection window of $25\mbox{ }\upmu \mbox{s}$ of illumination by pulse pairs arriving at $f_\mathrm{r}$ is chosen to minimize the probability of state mixing, for which the measured time constant is $69 \mbox{ }\upmu \mbox{s}$ \cite{SupplementalMaterials}. Detection of Rabi flopping of the qubit by the microwaves, shown in Fig.~\ref{fig:UnshelvedResults}b, shows a fringe visibility of $V\!=\!0.49 \pm 0.02$ when the phase of the interferometer delay is at the top of the desired fringe (magenta, also outlined in magenta at the top of Fig.~\ref{TriRamseyFig}b), while essentially no state discrimination is evident ($V\!=\!0.020\pm0.009$) at the bottom of the same fringe (black, outlined accordingly in \ref{TriRamseyFig}b). The photon arrival times in the desired phase configuration, shown in Fig.~\ref{fig:UnshelvedResults}a, show a clear distinction between the two qubit states, though the dark state is not as dark as was the case for the shelved qubit (compare to Fig.~\ref{ShelvedHistograms}a).  By taking our total detection efficiency into account, we estimate that the state dependence of the total number of emitted photons ($\approx\!5$, right vertical axis, Fig.~\ref{fig:UnshelvedResults}b) is about an order of magnitude less than $\kappa$.  The experimentally observed sensitivity of the state selectivity to the interferometer phase suggests that this may be due to drifts of the interferometer's path-length difference during measurements, which we are only able to monitor to tens of nanometers.  While the total number of detected photons per measurement is too low to allow single-shot measurement with this simple scheme, ensemble measurements allow clear state distinction, and coherent control techniques that are more sophisticated than this simple delay stage may be capable of allowing more cycling.

While the time-domain picture described above is sufficient for understanding the principles involved, the fact that we operate in the weak-pulse limit ($\theta \ll \pi$) suggests the existence of a simple frequency-domain interpretation.  In particular, it is tempting to view the Mach-Zehnder interferometer (Fig.~\ref{TriRamseyFig}d) as a \emph{frequency filter} that allows the desired frequencies to pass while the rejected frequencies are sent to the unused output port.  However, since the interferometer is implemented for the red (un-doubled) light, while the optical spectra exiting the two ports will be complimentary, they would produce almost indistinguishable spectra in the UV upon subsequent frequency doubling.  This is because the difference between the intra-pulse-pair phases of the two ports in the red is $\pi$, which becomes $2 \pi$ after frequency doubling. The frequency domain description that is probably most apt, then, is that the red light interferometer controls the creation of the UV light so that power spectral density near undesired frequencies is not created in the first place. While this frequency domain interpretation is perfectly applicable for this work, this subtlety of the filtration is one reason we have chosen to instead focus primarily on the time-domain picture here, which will likely continue to be applicable for stronger pulses.

For electron-shelved and optical-frequency ion qubits, the temporal filtering enabled by the use of a mode-locked laser for state detection has allowed us to achieve a state preparation and measurement fidelity approaching the highest reported observation with this atom \cite{Noek2013Optical} and should be compatible with future integrated high efficiency, non-imaging detectors.  Future development of ultrafast optical switches may someday allow extension of this idea to cw lasers \cite{Koulakis2019Sparks}. 
For direct hyperfine qubit detection, the simple delay scheme we have used here could be viewed as the minimal possible instance of coherent control, an approach that has been demonstrated in other contexts with far more sophisticated implementations.  Nonetheless, this simple version has been sufficient to tease out and control the dynamics of a quantum 4-level system with $\approx\! 30 \mbox{ MHz}$ precision and enable temporal filtering for qubit state detection.  The fact that useful dynamical information can be extracted from a single atom with such a simple system is largely due to the high level of control that exists for the isolated quantum systems employed for quantum information processing, and it seems likely that further application of the approaches of coherent control to the pure state systems in quantum information science (particularly for many-qubit systems) will lead to new an potentially useful insights.

\begin{acknowledgments}
The authors acknowledge Thomas Dellaert and Patrick McMillin for technical assistance, Adam West for graphical assistance, and E.~Hudson, P.~Hamilton, J.~Caram, and T.~Atallah for discussions.  This work was supported by the U.S.~Army Research Office under Grant No. W911NF-15-1-0261.

C. R. and A. R. contributed equally to this work.
\end{acknowledgments}

\bibliography{UltrafastStateDetection}
\end{document}